\documentclass[12pt]{iopart}

\usepackage{graphicx}
\usepackage{amsfonts}

\usepackage{epsfig}
\usepackage{graphicx}
\usepackage{dcolumn}
\usepackage{bm}
\usepackage{xcolor}

\begin{document}
	
	\title{Containing rumors spreading on correlated multiplex networks}
	
	\author{Jiajun Xian$^{1}$, Dan Yang$^{2}$, Liming Pan$^{2}$, Ming Liu$^{1}$, Wei Wang$^{3,2*}$}
	
	\address{$^{1}$ School of Computer Science and Engineering,
		University of Electronic Science and Technology of China, Chengdu, 611731, China}
	
	\address{$^{2}$ Web Sciences Center, School of Computer Science and Engineering,
		University of Electronic Science and Technology of China, Chengdu, 611731, China}
	
	\address{$^{3}$ Cybersecurity Research Institute, Sichuan University, Chengdu 610065, China}
	\address{E-mail: *wwzqbx@hotmail.com}

	
	\begin{abstract}
		Rumors flooding on rapidly-growing online social networks has geared much attention from many fronts. Individuals can transmit rumors via numerous channels since they can be active on multiple platforms. However, no systematic theoretical research of rumors containing dynamics on multiplex networks has been conducted yet. 
		In this study, we propose a family of containing strategies based on the degree product $\mathcal{K}$ of each user on the multiplex networks.
		Then, we develop a heterogeneous edge-based compartmental theory to comprehend the containing dynamics.
		The simulation results demonstrate that strategies with preference to block users with large $\mathcal{K}$ can significantly reduce the rumor outbreak size and enlarge the threshold. 
		Besides, better performance can be expected on heterogeneous multiplex networks with the increasing of preference intensity and degree heterogeneity. 
		Moreover, take the inter-layer degree correlations $r_s$ into consideration, 
		the strategy performs best on multiplex networks with $r_s=-1$, $r_s=1$ the second, and $r_s=0$ the last. 
		On the contrary, if we prefer to block users with small $\mathcal{K}$ rather than large $\mathcal{K}$, the containing performance will be worse than that of blocking users randomly on most multiplex networks except for uncorrelated multiplex networks with uniform degree distribution. We found that the blocking preferences have no influence on the containing results on uncorrelated multiplex networks with uniform degree distribution.
		Our theoretical analysis can well predict the rumors containing results and performance differences in all the cases studied.
		The systematic theoretical research of rumors containing dynamics on multiplex networks in this study will offer inspirations for further investigations on this issue.
	\end{abstract}

	
	\maketitle
	\tableofcontents
	\section{Introduction}
	
	Rumors are pieces of purportedly true information, of which the content can vary from traditional gossip to deliberate disinformation. Thus, the rumors can be considered to be a special case of misinformation~\cite{del2016spreading,pennycook2019fighting}.
	 In real life, the authenticity of a rumor is usually hard to confirm. It can be interpreted as a mental infection~\cite{nekovee2007theory}, which affects individuals' behaviors by shaking their opinions; thus, an elaborate rumor may cause serious hazards~\cite{sunstein2014rumors}, such as financial losses~\cite{galam2003modelling,kimmel2004rumors,kosfeld2005rumours}, public panic and social instability~\cite{moreno2004dynamics,zhang2009interplay,coast2015rumour}.
	Nowadays, the problem of rumor flooding has become even more severe than before with the emergence of online social platforms (e.g., Facebook, Twitter, and MySpace) since rumors can spread faster and have wide transmissions with the help of online networks~\cite{liu2019modeling}. Researchers in multiple fronts ( e.g., network science, applied mathematics, computer science, and other interdisciplinary disciplines) are making efforts to construct suitable rumor--spreading models, predict its outbreak size, and contain its spreading.   
	
	In the 1960s, Daley and Kendall first proposed the original rumor--spreading model, named Daley--Kendall (DK) model~\cite{daley1964epidemics,daley1965stochastic}. The DK model is also widely known as the Ignorant--Spreader--stifleR (ISR) model since the population was separated into three categories: ignorants (people who have never heard the rumor), spreaders (people who are spreading the rumor ), and stiflers (people who have heard the rumor but choose not to spread it). 
	Later, Maki and Thompson developed the Maki--Thompson (MK) model~\cite{maki1973mathematical} by modifying the DK model and investigated the spreading dynamics based on mathematical theory. 
	Afterward, a group of researchers took the nature of human psychology and sociology into consideration when developing the rumor--spreading models, such as the psychological motivation of spreading rumors~\cite{dang2016toward}, the forgetting and remembering mechanism~\cite{zhao2013rumor}, and herd mentality~\cite{wang2017drimux}. See more similar work in~\cite{afassinou2014analysis,hong2018sir,fazli2019dynamics,gu2008effect,zhao2012sihr,xia2015rumor}. Parallel to this, some researchers focus on revealing the importance of network structures to the spreading of rumor~\cite{lefevre1994distribution}. For example, Zenette~\cite{zanette2001critical,zanette2002dynamics} proved the existence of critical thresholds of the rumor spreading on small-world networks by performing a series of simulations. Then, Moreno~\cite{moreno2004dynamics} et al. extended the dynamic model to scale-free networks and discovered that degree heterogeneity would influence the spreading dynamics of rumors as well. Nekovee et al.~\cite{nekovee2007theory} further found that degree correlations could also have a great effect on the rumor spreading dynamics.  More similar studies can be found in~\cite{li2019rumor,roshani2012effects,ya2013rumor,wang2014siraru,wang2019containing}.  
	Note that, although the spreading of disease~\cite{wang2017unification} and rumor share some common features, it is impossible to capture their spreading dynamics by analogous models. There are major differences between the spreading mechanism of disease and rumor~\cite{daley1964epidemics,daley1965stochastic,li2019rumor}. For example, in epidemics, the recovery of the infected nodes is not related to others; they recover with a fixed probability. When it comes to the rumor spreading, however, the process by which spreaders become stiflers depends on the states of their neighbors.
	 
	All of the extensive studies mentioned above are addressing the problem of rumor spreading on single--layer networks, but few researchers have investigated this problem on multiplex networks up to now, although multiplex networks have been continually exploited in the past few years~\cite{pan2019optimizing,boccaletti2014structure,kivela2014multilayer,gao2012networks,pan2019optimal,wang2019impact,di2018multiple}. 
	Systematic theoretical research of rumor containing on multiplex networks is even more rare, in spite of the fact that it is critical to consider spreading models with multiplex networks~\cite{xian2019misinformation} since users can be active on multiple platforms. 
	This research gap may contribute to the difficulties in the characterization of the multiplex network~\cite{xia2019new,wang2014degree,wang2014asymmetrically} and the description of dynamic correlations between neighbors on the multiplex networks. 
	These mentioned difficulties make the analytical studying of the rumor spreading dynamics and containing strategy on multiplex networks more challenging than epidemics. Most recent related studies~\cite{eddine2019minimizing,han2018modeling} are weak in this regard and fail to give a precise theoretical prediction of rumor outbreak size and threshold. In view of this, we aim to propose rumor containing strategies on correlated multiplex networks, of which the containing dynamics can be well comprehended; thus, the containing results can be predicted.
	Inspired by the target immunization strategies~\cite{lu2016vital,wang2016statistical,cohen2003efficient,granell2013dynamical} proposed by researchers in the study of epidemics. In this study, we initially propose a family of containing strategies with different blocking preferences to contain rumors spreading on correlated multiplex networks, and develop a heterogeneous edge-based compartmental theory~\cite{miller2011edge,volz2008sir,miller2009spread,
		valdez2013temporal,valdez2012temporal,valdez2012intermittent,wang2018social} to comprehend the containing dynamics. 
	Subsequently, we verify the effectiveness of our containing strategies and compare their performance on multiplex networks with various topology structures and different inter-layer degree correlations. 
	
	In the remainder of this paper, we describe the rumor--spreading and rumor--containing process in Sec. \ref{sec:model} and introduce the theoretical analysis systematically in Sec. \ref{sec:theory}. Then, we present the simulations with respect to the theoretical predictions in Sec. \ref{sec:simulation}. Finally, we provide a conclusion in Sec. \ref{sec:conclusion}.
	
	\section{Model description} \label{sec:model}
	
	In this section, we will thoroughly introduce our model.
	We first describe the rumor--spreading process on multiplex networks and then propose a family of containing strategies.

	\subsection{Rumor spreading process}\label{sec:process}
	
\begin{figure}
		\centering
		\includegraphics[width=0.9\textwidth]{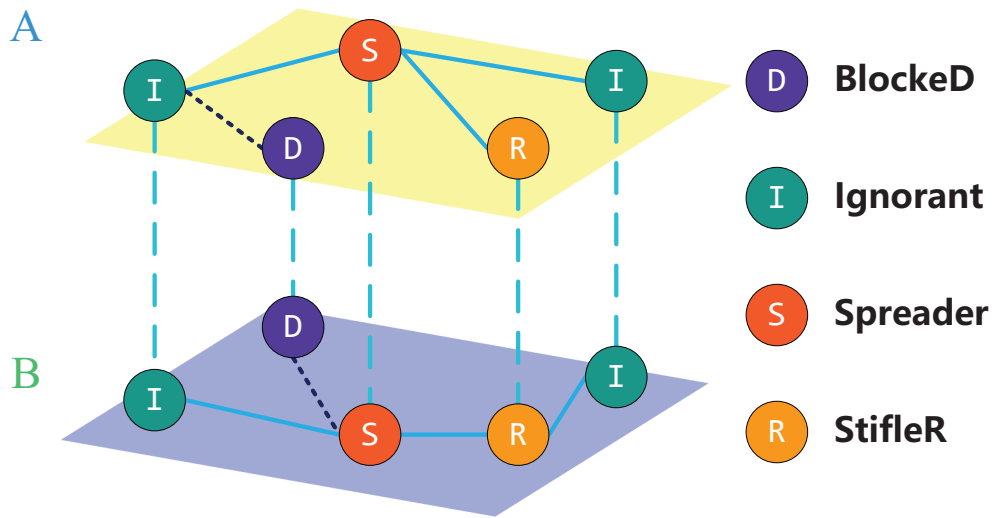}
		\caption{(Color online) Illustration of the rumor spreading model on multiplex networks. Nodes can be in four different states in the multiplex networks. Two one-to-one matched nodes in the distinct layer represent the same user and, thus, always have the same state.}\label{MISR}
	\end{figure}

	Social platforms can be abstracted into complex networks, where users (relations) can be represented by nodes (edges). In our model, rumors spread on a multiplex network with two layers: $A$ and $B$, as schematically illustrated in Fig. \ref{MISR}. Layers $A$ and $B$ have the same node set, representing the same user group. However, the two layers have distinct edge sets since the relations between users can be diverse in different platforms. In order to avoid confusion, for a user $u$, we denote the corresponding node in layer $x$ as $u_{x}$, where $x \in \{A,B\}$. Let $k_{x}$ be the degree of node $u_{x}$, then we use a vector $\vec{k}=(k_A, k_B)$ to denote the degree of a user who has $k_{A}$ $(k_{B})$ relations in layer $A$ $(B)$ for convenience. 
	Subsequently, the joint degree distribution of the multiplex network is given by $P(k_{A},k_{B})$, which is on behalf of the probability that a randomly selected user has degree $\vec{k}=(k_A, k_B)$. In real situations, the value of $k_{A}$ and $k_{B}$ can be correlated, and this kind of correlation is called the inter-layer degree correlation. Considering the deficiency of the Pearson coefficient in measuring the correlation of heterogeneous sequences~\cite{yang2017lower,guo2017bounds}, the Spearman rank correlation coefficient~\cite{lee2012correlated,cho2010correlated} is commonly used to quantify the underlined correlations, which is defined as
	\begin{equation}\label{eq:xiSIR}
	r_s=1-6\frac{\sum_{i=1}^N(d_{A}^{i}-d_{B}^{i})}{N(N^2-1)},
	\end{equation}
	where $d_{A}^{i}$ $(d_{B}^{i})$ is the degree ranking of node $u^{i}_{A}$ $(u^{i}_{B})$ in the layer $A$ $(B)$.
	The value of $r_{s}$ ranges in $[-1,1]$, where negative (positive) values indicate negative (positive) correlations and the larger absolute value, the stronger negative (positive) correlation. Previous studies have shown that the spreading dynamics on multiplex networks can be affected by the inter-layer degree correlations~\cite{wang2018social}; thus, we investigate the rumors spreading process on multiplex networks with different inter-layer degree correlations in this study. 
	For the simulation, we construct the multiplex networks by a generalized configuration model~\cite{catanzaro2005generation,wang2018social} for a given $P(k_{A},k_{B})$, which ensures that each layer of the multiplex networks is free of self-loops and multiple edges. The configuration model also stipulates that the inner-layer degree correlations can be neglected when the multiplex networks are vast and sparse, while the inter-layer degree correlations exist in our model and can be captured by $P(k_{A},k_{B})$.  
	
	The rumors spreading model we propose in this study can be described as an extension of the Ignorant-Spreader-stifleR model ~\cite{daley1964epidemics,daley1965stochastic}. As shown in Fig. \ref{MISR}, each node in the two layers can be in four distinct states, that is, 
	(1) blocked state (D), in which the nodes represent users who are blocked from accessing the rumor during the spreading; thus, they cannot accept or spread the rumor;
	(2) ignorant state (I), where the nodes stand for users who are unaware of the rumor; (3) spreader state (S), in which the nodes represent users who have accepted the rumor and are ready to spread it; (4) stifler state (R), where the nodes stand for users who used to be spreaders but have lost interest in the rumor. Therefore, we denote our model as the DISR model for convenience. Notice that the blocked nodes are strategically selected before the rumor spreading begins to contain its spreading. 
	A detailed strategy description will be found in Sec. \ref{sec:strategy}.
	After that, we initiate the spreading by randomly selecting a user to be a spreader; thus, the corresponding node in both layer $A$ and $B$ should be set to the S state. Actually, each node in layer $A$ and its counterpart in layer $B$ share the same state since they represent the same user. For every time step, spreader nodes transmit the rumor to their neighbors, but only those nodes in the state I will accept the rumor and become spreaders with a probability of $\lambda$ in the next time step. After all the spreaders finish trying to transmit the rumor, each of them turns into the R state with a probability of $1-(1-\gamma)^{n}$, where $n$ is the total number of neighbors in the S or R state that the spreader individual has in either layer $A$ or $B$, and $\gamma$ denotes the unit probability when $n=1$. The rumor spreading will be terminated once there is no spreader in the multiplex networks.
	
	\subsection{Rumor containing strategy }\label{sec:strategy}
	
	To contain the rumor spreading on multiplex networks, we should develop effective strategies. In this study, we present a family of containing strategies dependent on the degree of nodes in both layers. The main idea of the containing strategy is to block a fraction of $\rho$ of users with reasonable preferences before the spreading begins.  
	Firstly, we assign each user $i$ a weight value $W_{\delta}(k_A^{i},k_B^{i})$, which represents the probability that user $i$ is selected to be blocked, and the corresponding nodes $u^{i}_{A}$ and $u^{i}_{B}$ are set to be in the D state before the spreading process begins.
	Afterward, we select a fraction of $\rho$ of users according to their assigned values.
	The exact calculation of the value $W_{\delta}(k_A^{i},k_B^{i})$ is as follows,
	\begin{equation}\label{eq:www}
	W_{\delta}(k_A^{i},k_B^{i})=\frac{(k_A^{i}k_B^{i})^{\delta}}{\sum^{N}_{i=1}(k_A^{i}k_B^{i})^{\delta}},     -\infty<\delta<+\infty
	\end{equation} 
	where $k_A^{i} (k_B^{i})$ is the degree of node $u^{i}_{A}(u^{i}_{B})$, and the different values of $\delta$ indicate different blocking preferences. Denote $\mathcal{K} = k_A^{i}k_B^{i}$, in the case $\delta > 0$ ($\delta < 0$), user $i$ with larger $\mathcal{K}$ will have more (less) probability to be blocked. 
	In a particular case, when $\delta = 0$, we randomly choose users to block.  
	The effects of different blocking preferences will be compared in detail in Sec. \ref{sec:simulation}.
	
	\section{Theoretical analysis} \label{sec:theory}
	
	In this section, we will present our theoretical analysis of the spreading process and containing strategies in detail. 
	To begin with, we obtain the probability for each user to be blocked after the blocking process according to the containing strategy described in Sec. \ref{sec:strategy}. Subsequently, inspired by Refs.~\cite{miller2011edge,volz2008sir,miller2009spread,
		valdez2013temporal,valdez2012temporal,valdez2012intermittent,wang2018social}, we develop a heterogeneous edge-based compartmental theory to comprehend the containing dynamics. We focus on the analysis of rumor outbreak size $R(\infty)$ versus effective transmission probability $\beta=\lambda/\gamma$.
	Additionally, we obtain the rumor outbreak threshold $\beta_c$, above which the rumor will break out, by a stability analysis based method.
	
	\subsection{Containing strategies analysis} \label{sec:strategy analysis}
	
	We aim to find the probability that a user with degree $\vec{k}=(k_A,k_B)$ becomes blocked by our blocking strategy in this subsection.
	After blocking a fraction $\rho$ of users according to Eq. (\ref{eq:www}), there will be a fraction of $q=1-\rho$ users remain. Let $N_q(\vec{k})$ be the number of users with degree $\vec{k}=(k_A,k_B)$ in the remaining users, then the joint degree distribution $P_q(k_A,k_B)$ of the remaining users would be
	\begin{equation}\label{eq:pp}
	P_q(\vec{k})=\frac{N_q(\vec{k})}{qN}.
	\end{equation}
	When another user is blocked, $N_q(\vec{k})$ changes as
	\begin{equation}\label{eq:ap}
	N_{(q-1/N)}(\vec{k})=N_q(\vec{k})-\frac{P_q(\vec{k})(k_Ak_B)^{\delta}}{\langle (k_Ak_B)^{\delta}(q) \rangle},
	\end{equation}
	where $\langle (k_Ak_B)^{\delta}(q) \rangle \equiv \sum_{\vec{k}} P_q(\vec{k})(k_Ak_B)^{\delta}$. In the limit of $N\rightarrow\infty$, Eq. (\ref{eq:ap}) can be represented in terms of derivatives of $N_{(q-1/N)}(\vec{k})$ with respect to $q$, that is,
	\begin{equation}\label{eq:dap}
	\frac{dN_q(\vec{k})}{dq}=N\frac{P_q(\vec{k})(k_Ak_B)^{\delta}}{\langle (k_Ak_B)^{\delta}(q) \rangle}.
	\end{equation}
	Differentiating Eq. \ref{eq:pp} with respect to $q$ and using Eq. (\ref{eq:dap}), we obtain
	\begin{equation}\label{eq:pdp}
	-q\frac{dP_q(\vec{k})}{dq}=P_q(\vec{k})-\frac{P_q(\vec{k})(k_Ak_B)^{\delta}}{\langle (k_Ak_B)^{\delta}(q) \rangle},
	\end{equation}
	which is exact for $N\rightarrow\infty$. In order to solve Eq. (\ref{eq:pdp}), we define the function $G_\delta(x) \equiv \sum_{\vec{k}} P(\vec{k})x^{(k_1k_2)^{\delta}}$, which is absolutely convergent and monotonically increases for $x \in [0, 1]$, and $G_\delta(0)=0$ and $G_\delta(1)=1$.  Inspired by Refs.~\cite{huang2011robustness,shao2009structure}, we introduce a new variable, $t \equiv G_{\delta}^{-1}(q)$, and then, by direct differentiation, we obtain that
	\begin{equation}\label{eq:pp2}
	P_q(\vec{k})=P(\vec{k})\frac{t^{(k_Ak_B)^{\delta}}}{G_{\delta}(t)}=\frac{1}{q}P(\vec{k})t^{(k_Ak_B)^{\delta}}
	\end{equation}
	satisfies Eq. (\ref{eq:pdp}). Thus, the probability that a user with degree $\vec{k}=(k_{A}, k_{B})$ become blocked when the blocking fraction is set to be $\rho$ should be
	\begin{equation}\label{eq:pm}
	p_q^{D}(\vec{k})=\frac{NP(\vec{k})-pNP_p(\vec{k})}{NP(\vec{k})}=1-t^{(k_1k_2)^{\delta}}.
	\end{equation}
	Accordingly, the nodes that represent users of degree $\vec{k}=(k_{A}, k_{B})$ are set to be in the D state with a probability of $p_q^{D}(\vec{k})$ before the spreading begins.
	
	\subsection{Containing dynamics of the rumor}\label{sec: spreading dynamics}
	In order to comprehend the rumor dynamics, we use a developed heterogeneous edge-based compartmental theory in this subsection. For simplicity, nodes of identical degrees are assumed to have the same dynamic properties in our analysis.
	Let $\zeta_{x}(\vec{k},t)$ denote the probability that node $v_{x}$ has not transmitted the rumor to neighboring node $u_{x}$ by time $t$, where $\vec{k}=(k_A,k_B)$ is the degree of the corresponding user of $v_{x}$.
	As described in our model, nodes can be in four different states; thus, we further divide $\zeta_{x}(\vec{k},t)$ as
	\begin{equation}\label{eq:xiSIR}
	\zeta_{x}(\vec{k},t)=\eta_{x}^{I}(\vec{k},t)+\eta_{x}^{S}(\vec{k},t)+\eta_{x}^{R}(\vec{k},t)+\eta_{x}^{D}(\vec{k},t),
	\end{equation}
	where the value of $\eta_{x}^{I}(\vec{k},t)$, $\eta_{x}^{S}(\vec{k},t)$, $\eta_{x}^{R}(\vec{k},t)$ and $\eta_{x}^{D}(\vec{k},t)$ denote
	the probability of $v_{x}$ being in the I, S, R and D state, respectively, and has not transmitted rumor to $u_{x}$ up to time $t$. 
	Notice that for a given blocking fraction $\rho$, $\eta_{x}^{D}(\vec{k},t)$ should be a constant for each node $v_{x}$ in the spreading process since the blocking process finishes before the spreading begins. Through direct derivation, we can get 
	\begin{equation}\label{eq:xiM}
	\eta_{x}^{D}(\vec{k},t)=p_q^{D}(\vec{k}),
	\end{equation}
	where $p_q^{D}(\vec{k})$ is obtained by Eq. (\ref{eq:pm}).
	
	For the dynamic correlations among the rest of the states of nodes, we refer to the cavity theory and assume $u_x$ to be in the cavity state, where $u_x$ can receive rumors but cannot transmit them. According to the model, ignorant nodes will turn into the spreader state with a probability of $\lambda$ when they get contacted by a spreader neighbor; thus, the evolution of $\zeta_{x}(\vec{k},t)$ are strongly correlated with $\eta_{x}^{S}(\vec{k},t)$, to be exact,
	\begin{equation}\label{eq:xiS}
	\frac{d\zeta_{x}(\vec{k},t)}{dt}=-\lambda\eta_{x}^{S}(\vec{k},t).
	\end{equation}
	Besides, nodes in the S state will turn to the R state with a certain probability; thus, the growth of $\eta_{x}^{R}(\vec{k},t)$ should be
	\begin{equation}\label{eq:xiR}
	\frac{d\eta_{x}^{R}(\vec{k},t)}{dt}=(1-\lambda)\eta_{x}^{S}(\vec{k},t)[1-(1-\lambda)^{n_{x}(\vec{k},t)}],
	\end{equation}
	where $n_{x}(\vec{k},t)$ is the average number of neighbors in the S or R state that the spreader user might have in either layer A or B; the detailed calculation of $n_{x}(\vec{k},t)$ will be presented in~\ref{nx}.
	We further consider the average probability that the rumor has not been transmitted to node $u_x$ by $v_x$,
	\begin{equation}\label{eq:theta}
	\zeta_{x}(t)=\frac{1}{\langle k \rangle_x}\sum_{\vec{k}} k_{x} P(\vec{k})\zeta_{x}(\vec{k},t),
	\end{equation} 
	where $\langle \cdot \rangle_x$ denotes the average degree of layer $x$.
	In layer $x$, with the assumption that $u_{x}$ is in the cavity state, $v_{x}$ can only receive rumor from $(k_{x}-1)$ neighbors; thus, 
	\begin{equation}\label{eq:xiI}
	\eta_{x}^{I}(\vec{k},t)=[1-p_q^{D}(\vec{k})]\zeta_{x}(t)^{k_{x}-1}\zeta_{y}(t)^{k_{y}},
	\end{equation}
	where $y \in \{A, B\}$ and $y\neq x$ denotes the counterpart of $x$, 
	
	Combining Eqs. (\ref{eq:xiSIR})-(\ref{eq:xiI}), we get the probability that a user with degree $\vec{k}=(k_{A},k_{B})$ is ignorant up to time $t$ as
	\begin{equation}\label{eq:Ik}
	p^I_{q}(\vec{k},t)=[1-p_q^{D}(\vec{k})]\zeta_{x}(t)^{k_{A}}\zeta_{y}(t)^{k_{B}}.
	\end{equation}
	Therefore, we obtain the density of ignorant
	users at time $t$ as
	\begin{equation}\label{eq:I}
	I(t)=\sum_{\vec{k}}P(\vec{k})p^I_{q}(\vec{k},t).
	\end{equation}
	When the spreading process is terminated, that is, $t\rightarrow\infty$, users can only be in the D, I or R state; thus, the rumor outbreak size should be
	\begin{equation}\label{eq:I}
	R(\infty)=q-I(\infty).
	\end{equation}
	
	\begin{figure}
		\centering
		\includegraphics[width=0.9\textwidth]{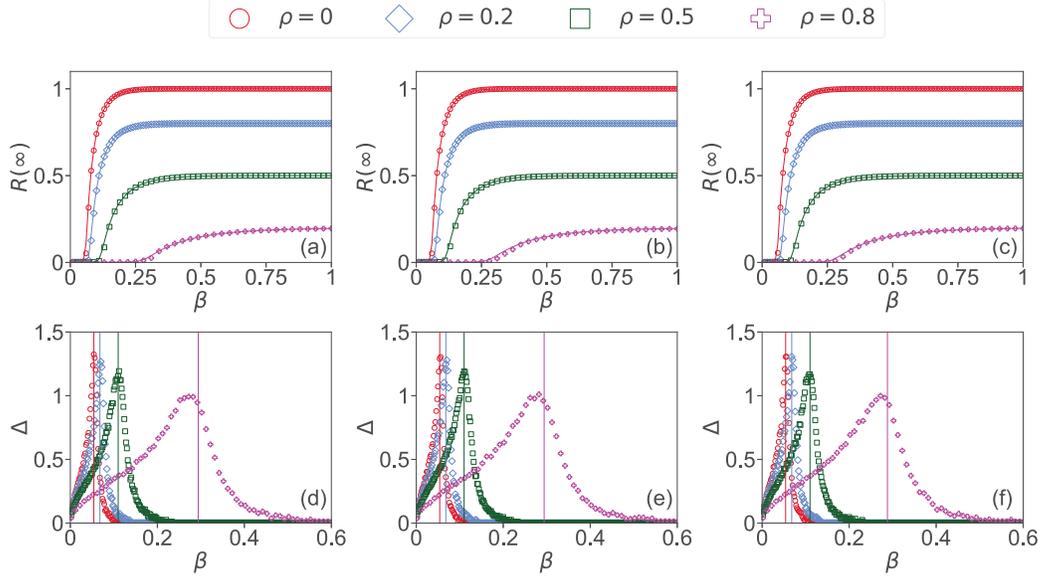}
		\caption{(Color online) Strategy performance on uncorrelated ER-ER multiplex networks with average degree $\left\langle k_A\right\rangle=\left\langle k_B\right\rangle=10$. The final break size $R(\infty)$ versus $\beta$ after blocking a fraction $\rho$ of users with blocking preference (a) $\delta=-5$, (b) $\delta=0$ and (c) $\delta=5$. The variability $\Delta$ of $R(\infty)$ versus $\beta$ after blocking a fraction $\rho$ of users with blocking preference (d) $\delta=-5$, (e) $\delta=0$ and (f) $\delta=5$. The simulation results when $\rho=0$, $\rho=0.2$, $\rho=0.5$ and $\rho=0.8$ are denoted by red circles, blue diamonds, green squares, and magenta pluses, respectively, and the corresponding theoretical predictions are denoted by solid lines. Vertical lines inside the plot of (d)-(c) mark the positions of the corresponding rumor outbreak thresholds predicted by our theory. }\label{R_beta}
	\end{figure}
	\begin{figure}
		\centering
		\includegraphics[width=0.9\textwidth]{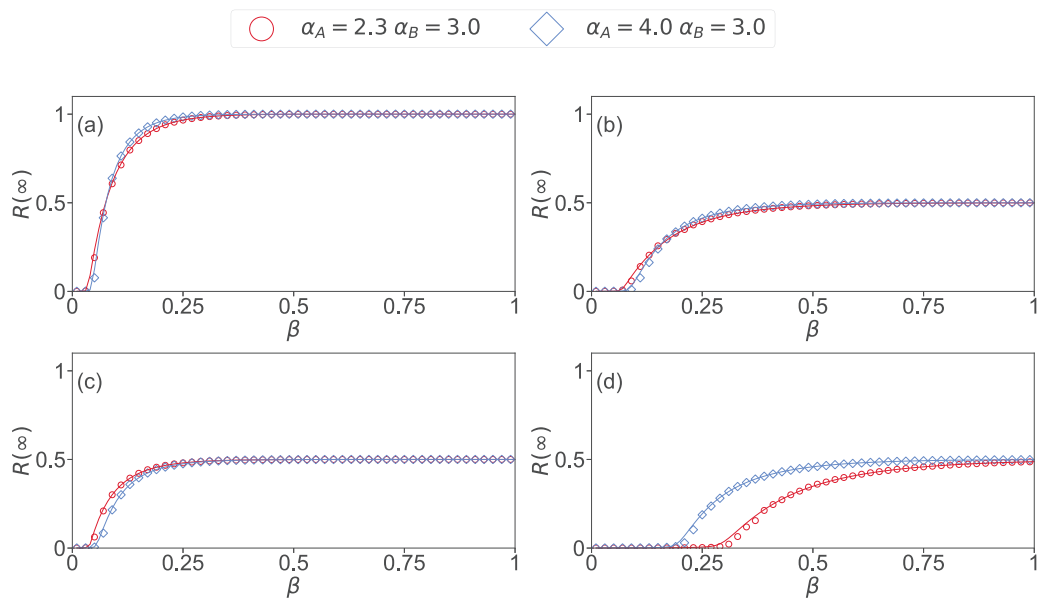}
		\caption{(Color online) Strategy performance on uncorrelated SF-SF multiplex networks. The final break size $R(\infty)$ versus $\beta$ with (a) no containing process, and after blocking 50\% users with blocking preference (b) $\delta=0$, (c) $\delta=-5$ and (d) $\delta=5$. The simulation results on SF-SF multiplex networks with degree exponent $\alpha_A=2.3$ and $\alpha_B=3.0$ ($\alpha_A=3.0$ and $\alpha_B=4.0$) are denoted by red circles (blue diamonds), and the corresponding theoretical predictions are denoted by solid lines.  
		}\label{hetero}
	\end{figure}
	\begin{figure}
		\centering
		\includegraphics[width=0.9\textwidth]{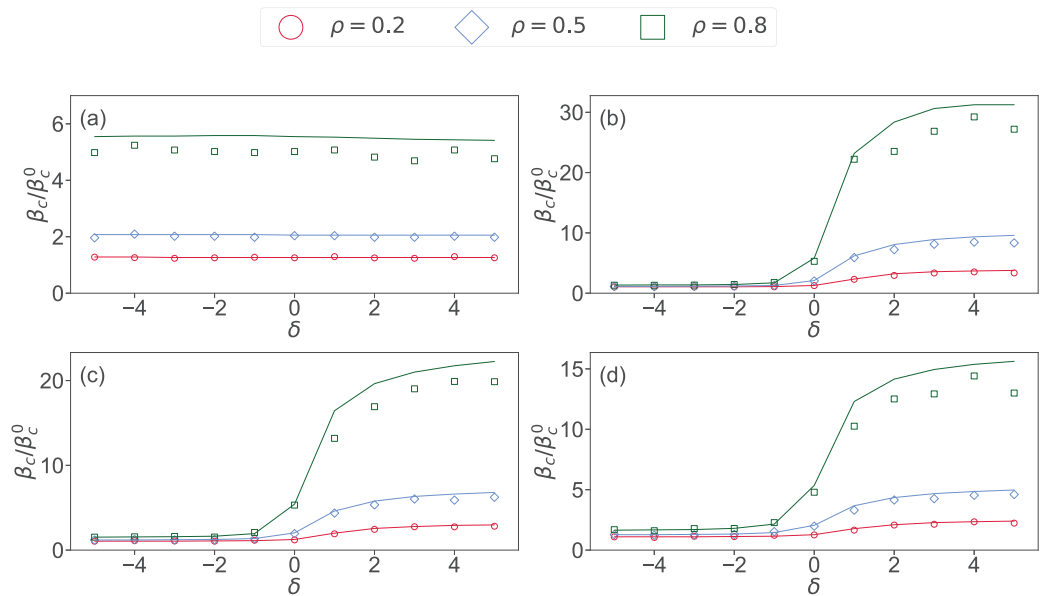}
		\caption{(Color online) The effectiveness of containing strategy in enlarging the rumor outbreak threshold. $\beta_c/\beta_c^{0}$ versus $\delta$ on (a) $ER-ER$ multiplex networks with $\left\langle k_A\right\rangle=\left\langle k_B\right\rangle=10$; (b) $SF-SF$ multiplex networks with $\alpha_A=2.3$ and $\alpha_B=3.0$; (c) $SF-SF$ multiplex networks with $\alpha_A=\alpha_B=3.0$; and (d) $SF-SF$ multiplex networks with $\alpha_A=4.0$ and $\alpha_B=3.0$. The average degree of $SF-SF$ multiplex networks are all set as $\left\langle k_A\right\rangle=\left\langle k_B\right\rangle=10$. The red circles, blue diamonds, and green squares denote the simulation results when $\rho=0.2$, $\rho=0.5$, and $\rho=0.8$, respectively. The corresponding theoretical predictions are denoted by solid lines.   }\label{betabeta}
	\end{figure}
	\begin{figure}
		\centering
		\includegraphics[width=0.9\textwidth]{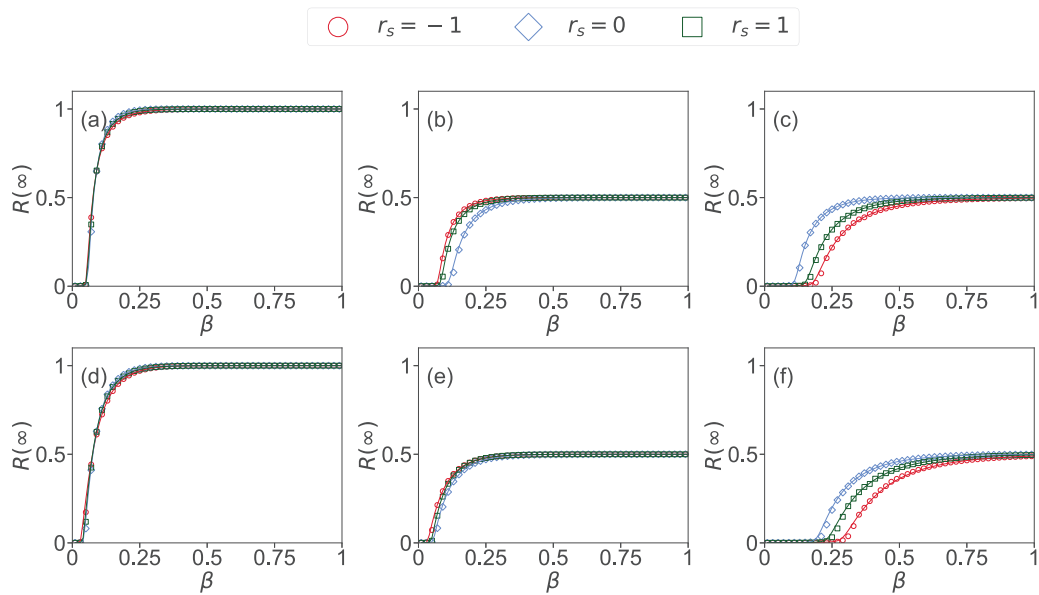}
		\caption{(Color online) Strategy performance on multiplex networks with different inter-layer degree correlations. The final break size $R(\infty)$ versus $\beta$ on ER-ER multiplex networks with $\left\langle k_A\right\rangle=\left\langle k_B\right\rangle=10$, after blocking (a) no user, or blocking 50\% users with blocking preference (b) $\delta=-5$, and (c) $\delta=5$. The final break size $R(\infty)$ versus $\beta$ on SF-SF multiplex networks with $\alpha_A=\alpha_B=3.0$, after blocking (d) no user, or blocking 50\% users with blocking preference (e) $\delta=-5$, and (f) $\delta=5$. 
			The red circles, blue diamonds, green squares denote the simulation results when $r_s=-1$, $r_s=0$, and $r_s=1$, respectively. The corresponding theoretical predictions are denoted by solid lines. }\label{rms}
	\end{figure}
	\begin{figure}
		\centering
		\includegraphics[width=0.9\textwidth]{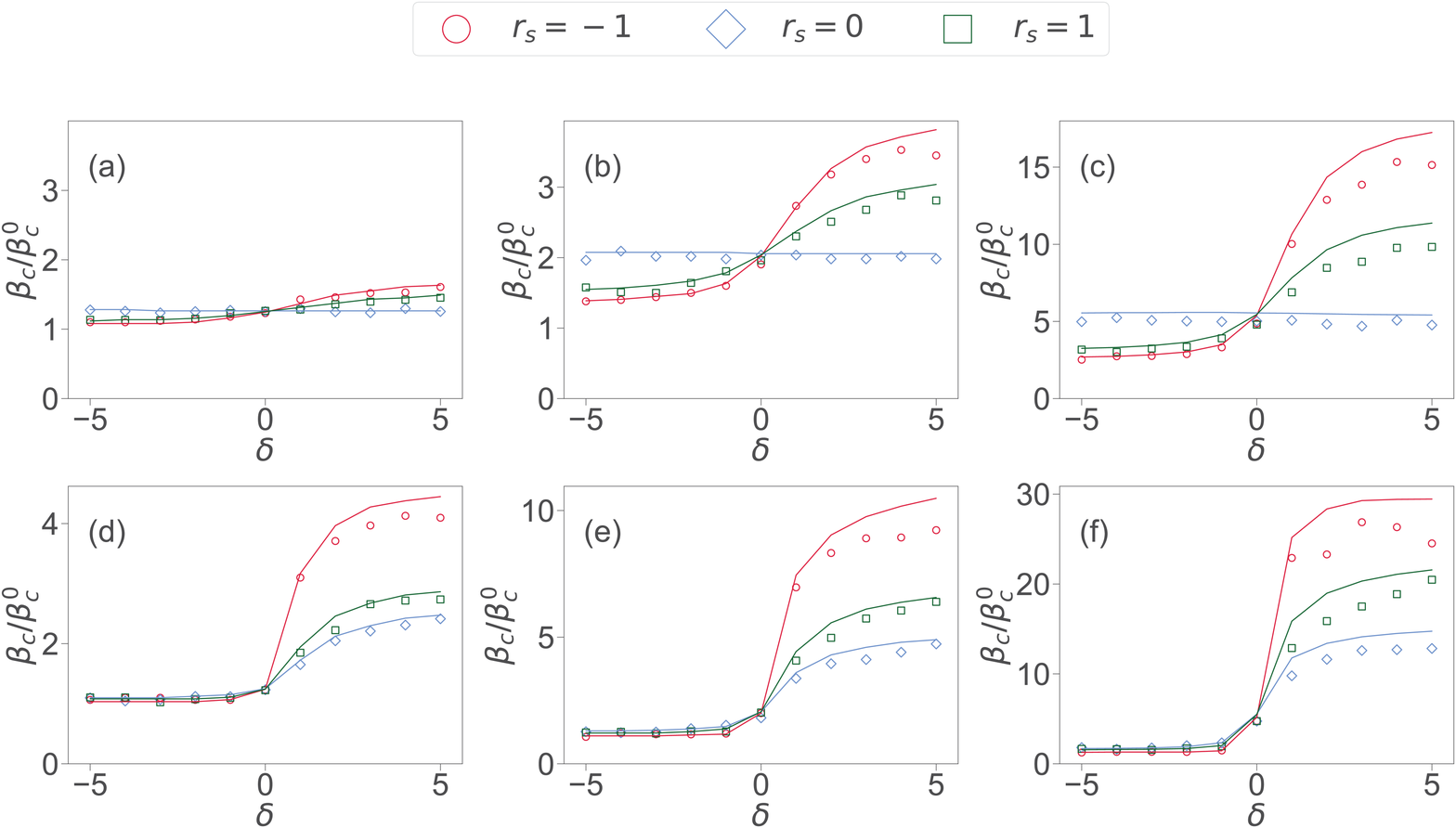}
		\caption{The effectiveness of containing strategy in enlarging the rumor outbreak threshold on multiplex networks with different inter-layer degree correlations. $\beta_c/\beta_c^{0}$ versus $\delta$ on ER-ER multiplex networks with (a) $\rho=0.2$, (b) $\rho=0.5$, and (c) $\rho=0.8$. And $\beta_c/\beta_c^{0}$ versus $\delta$ on SF-SF multiplex networks with (d) $\rho=0.2$, (e) $\rho=0.5$, and (f) $\rho=0.8$. The average degrees of the ER-ER and SF-SF multiplex networks are set as $\left\langle k_A\right\rangle=\left\langle k_B\right\rangle=10$, and the degree exponents of SF-SF multiplex networks are $\alpha_A=\alpha_B=3.0$.
			The red circles, blue diamonds, green squares denote the simulation results when $r_s=-1$, $r_s=0$, $r_s=1$, respectively, and the corresponding theoretical predictions are denoted by solid lines.}\label{betams}
	\end{figure}

	\subsection{Threshold analysis}\label{sec:threshold}
	In this subsection, we use a stability analysis based method to obtain the rumor outbreak threshold $\beta_c$ on the basis of dynamical equations acquired in Sec. \ref{sec: spreading dynamics}.  
	According to Eq. (\ref{eq:xiSIR}), there are five group of dynamical variables, that is, $\zeta_{x}(\vec{k},t)$, $\eta_{x}^{I}(\vec{k},t)$, $\eta_{x}^{S}(\vec{k},t)$, $\eta_{x}^{R}(\vec{k},t)$, and $\eta_{x}^{D}(\vec{k},t)$ for $x\in \{A,B\}$.
	Note that $\eta_{x}^{D}(\vec{k},t)$ is a constant in the spreading process. Then, using the relations Eqs. (\ref{eq:xiSIR}) and (\ref{eq:xiI}), $\eta_{x}^{S}(\vec{k},t)$ and $\eta_{x}^{I}(\vec{k},t)$ can be eliminated; thus, we get the remaining relations
	\begin{equation}\label{eq:thresholdEquation}
	\begin{large}
	\left\{\begin{array}{l}\frac{d\zeta_{x}(\vec{k},t)}{dt}=\rho_{x}(\vec{k},t),\\
	\\\frac{ d\eta_{x}^{R}(\vec{k},t)}{dt}= \sigma_{x}(\vec{k},t),\end{array}\right.
	\end{large}
	\end{equation}
	where 
	\begin{equation}
	\rho_{x}(\vec{k},t)=-\lambda \left(\zeta_{x}(\vec{k},t)-\eta_{x}^{R}(\vec{k},t)-\eta_{x}^{I}(\vec{k},t)-\eta_{x}^{D}(\vec{k},t)\right)
	\end{equation}
	and
	\begin{equation}\label{}
	\sigma_{x}(\vec{k},t)=\left(1-\lambda^{-1}\right)[1-(1-\lambda)^{n_{x}(\vec{k},t)}]\rho_{x}(\vec{k},t).
	\end{equation}
	For Eq.(\ref{eq:thresholdEquation}), there is always a trivial fixed point $\zeta_{x}(\vec{k},t)=p_q^{D}(\vec{k})$ and $\eta_{x}^{R}(\vec{k},t)=0$, since $\eta_{x}^{D}(\vec{k},t)$ is a constant. Denote the Jacobian matrix of the dynamical system as $\mathcal J$, inspired by the approach in Ref. ~\cite{xian2019misinformation}, we can obtain the rumor outbreak threshold $\beta_c$ by checking the leading eigenvalue $\nu$ of $\mathcal J$ at the given fixed point. Let $\mathcal \tau$ to denote the number of distinct degree $\vec{k}$ appearing in the multiplex network. Through some careful calculations, we can get the Jacobian matrix $\mathcal J$ at the underlined point as
	\begin{equation}
	\mathcal J=
	\begin{footnotesize}
	\left(
	\begin{array}{cccc}
	\lambda (\Theta_A-I)&\lambda \Omega_{AB}&\lambda I&0\\
	\lambda \Omega_{BA}&\lambda (\Theta_B-I)&0&\lambda I\\
	\gamma\left(\lambda-1\right) (\Theta_A-I)&\gamma\left(\lambda-1\right)\Omega_{AB}& \gamma\left(\lambda-1\right)I&0\\
	\gamma\left(\lambda-1\right)\Omega_{BA}&\gamma\left(\lambda-1\right) (\Theta_B-I)&0& \gamma\left(\lambda-1\right)I \\
	\end{array}
	\right),
	\end{footnotesize}
	\end{equation}
	where $\Theta_x$ is a $\mathcal \tau \times \mathcal \tau$ matrix indexed by $\vec{k}$ with elements
	\begin{equation}
	\Theta^{\vec{k},\vec{k^{\prime}}}_x=[1-p_q^{D}(\vec{k})]\frac{k_{x}^{\prime}(k_{x}-1)P(\vec{k^{\prime}})}{\langle k\rangle_x},
	\end{equation}
	and $\Omega^{xy}$ is a $\mathcal \tau \times \mathcal \tau$ matrix  with elements
	\begin{equation}
	\Omega^{\vec{k},\vec{k^{\prime}}}_{xy}=[1-p_q^{D}(\vec{k})]\frac{k_{y}k_{y}^{\prime}P(\vec{k^{\prime}})}{\langle k\rangle_y}.
	\end{equation}
	As demonstrated in Ref.~\cite{xian2019misinformation}, when $\beta<\beta_c$, $\nu$ will stay close to zero, but once $\beta$ exceeds $\beta_c$, $\nu$ will deviate from zero and expand with $\beta$; thus, the value of $\beta_c$ can be determined by checking the value of $\nu$ versus $\beta$. 
	
	\section{Simulation results}\label{sec:simulation}
	
	In this section, we test the effectiveness of our containing strategies and the accuracy of our theoretical predictions by performing extensive numerical simulations on representative artificial multiplex networks of different types. Furthermore, we investigate the performance difference of our strategies on multiplex networks with different network topology and inter-layer degree correlation.
	Before the spreading begins, we select a fraction of $\rho$ of users to be in the D state, according to Eq. (\ref{eq:www}), and then initial the spreading by randomly choosing one remaining user to be a spreader. 
	All of the simulation results are averaged over 1000 independent realizations on a fixed multiplex network, of which each layer has 10000 nodes.
	
	First, we consider the uncorrelated ER-ER multiplex networks, each layer of which is an Erd\"{o}s-R\'{e}nyi (ER) network and there is no inter-layer degree correlation. Figs. \ref{R_beta}(a)-(c) show the corresponding results of outbreak size $R(\infty)$ versus effective transmission probability $\beta$ with different blocking fractions $\rho$ when $\delta = 5$, $\delta = 0$ and $\delta = -5$, respectively. For all the situations investigated, blocking a certain fraction of users can effectively reduce $R(\infty)$ and enlarge $\beta_c$, inhibiting the rumor spreading. Besides, we use a numerical method in Ref.~\cite{shu2018social} to obtain the values of $\beta_c$ by finding the peak of variability $\Delta$ of $R(\infty)$, where
	\begin{equation}
	\Delta=\frac{\sqrt{\left\langle R(\infty)^2\right\rangle-\left\langle R(\infty)\right\rangle^2}}{\left\langle R(\infty)\right\rangle}.
	\end{equation}
	Results shown in Figs. \ref{R_beta}(d)-(f) illustrate that increasing the blocking fraction will significantly enlarge the $\beta_c$. It can be seen that our theoretical predictions of $R(\infty)$ and $\beta_c$ coincide with the simulation results.
	
	Secondly, we port the spreading process on uncorrelated SF-SF multiplex networks, of which each layer is a scale-free (SF) network, as the degree distribution of most real-world networks follows a power-law distribution, that is, $p(k) \sim k^{-\alpha}$, where $\alpha$ denotes the degree exponent. More heterogeneous SF networks will have $\alpha$ of smaller values.  
	Two groups of SF-SF multiplex networks are employed in our simulation. The degree exponents of layers $A$ and $B$ of the multiplex networks in the first group are set as $\alpha_A = 2.3$ and $\alpha_B = 3.0$, respectively. In the second group, we set $\alpha_A = 3.0$ and $\alpha_B = 4.0$; thus, the multiplex networks in the second group are less heterogeneous than those in the first group.
	Comparing the containing results when $\delta = -5$ and  $\delta = 5$ in Figs. \ref{hetero} (c) and (d), respectively, one will see that the strategy with preference to block users with large $\mathcal{K}$ is much more effective than blocking users with small $\mathcal{K}$. We further compare the results in Figs. \ref{hetero} (b) and (c), and find that the strategy of blocking users with small $\mathcal{K}$ performs even worse than blocking users randomly in rumor containing.
	This can be explained, according to Eq. (\ref{eq:Ik}), which indicates that users with more relations have a higher average probability of accepting the rumor and then transmitting it. 
	Fig. \ref{hetero} (a) presents the spreading results on the two groups of multiplex networks with no containing process conducted. The results demonstrate that rumors spreading on more heterogeneous multiplex networks will have a larger (smaller) $R(\infty)$ for the small (large) value of $\beta$. Besides, comparing the results in Figs. \ref{hetero} (a) and (b), one can see that this kind of relative relationship between the results of $R(\infty)$ in the two groups of multiplex networks will not be changed by randomly selecting a fraction of $\rho$ users to block. However, when $\delta = -5$ and $\delta = 5$, as is showed in Figs. \ref{hetero} (c) and (d), respectively, the relative relationships between results is different from that in Fig. \ref{hetero} (a). To be specific, the strategy of blocking users with small (large) $\mathcal{K}$ will perform better in the less (more) heterogeneous multiplex networks. 
	This can be explained qualitatively, inferring to the fact that users with large $\mathcal{K}$ can facilitate the spreading.  
	When $\delta = -5$, more hub users with very large $\mathcal{K}$ will remain in the more heterogeneous multiplex networks; thus, the rumor is easier to break out. On the contrary, when $\delta = 5$, hub users are preferred to be blocked in the more heterogeneous multiplex networks, and a large number of users with very small $\mathcal{K}$ will remain, the spreading is not easy to break out.
	Again, the simulation results agree well with our theoretical predictions.
	
	We further investigate $\beta_c/\beta_c^{0}$ versus $\delta$ in different situations, where $\beta_c^{0}$ denotes the rumor outbreak threshold when there is no containing process. Larger $\beta_c/\beta_c^{0}$ indicates better containing performance. As shown in Fig. \ref{betabeta} (a), for uncorrelated ER-ER multiplex networks, given the blocking fraction, the value of $\beta_c/\beta_c^{0}$ stays the same for any value of $\delta$. This can be explained by the uniform degree distribution of ER networks. For uncorrelated SF-SF multiplex networks with heterogeneous degree distribution, as illustrated in Figs. \ref{betabeta} (b)-(d), only containing strategies with $\delta>0$ will significantly enlarge the rumor outbreak threshold. Larger $\delta$ will bring better performance. Besides, for any given $\delta>0$, larger $\beta_c/\beta_c^{0}$ can be expected on multiplex networks with higher heterogeneity. It can be seen that all these performing differences can be well comprehended by our theory. 
	
	Finally, we would like to investigate the influence of inter-layer degree correlations on the rumor containing dynamics. As shown in Fig. \ref{rms} (a), when there is no containing strategy conducted, rumor spreading on ER-ER multiplex networks with the same layer structures but different inter-layer degree correlations will have a similar growth pattern of $R(\infty)$.  
	Figs. \ref{rms} (b)-(c) present the results of $R(\infty)$ when $\delta = -5$ and $\delta = 5$, respectively. 
	According to the result, containing strategy with $\delta = -5$ ($\delta = 5$) works best in the multiplex networks with $r_s=0$ ($r_s=-1$), because when we prefer to block the users with small (large) $\mathcal{K}$, there will be less users with large $\mathcal{K}$ remain in the ER-ER multiplex networks with $r_s=0$ ($r_s=-1$). Besides, our strategy always gets the medium performance in multiplex networks with $r_s=1$ among the three groups of multiplex networks, regardless of the blocking preference. 
	This can also be confirmed by investigating the difference of $\beta_c/\beta_c^{0}$ in different situations.
	Figs. \ref{betams} (a)-(c) present the results of $\beta_c/\beta_c^{0}$ versus $\delta$ on ER-ER multiplex networks with different inter-layer degree correlations when the blocking fraction is set as $f=0.2$, $f=0.5$ and $f=0.8$, respectively.  The results reveal that, when $\delta>0$, on the multiplex networks with $r_s=-1$ the strategy can obtain the best performance, $r_s=1$ the second, and $r_s=0$ the last.
	On the contrary, if $\delta<0$, the strategy will be much less effective in all the cases studied. When it comes to the SF-SF multiplex networks, we obtain the same conclusion about $R(\infty)$ and $\beta_c/\beta_c^{0}$ as shown in Figs. \ref{rms} (d)-(f) and Figs. \ref{betams} (d)-(f), respectively. It can be seen that our theoretical predictions coincide with all the simulation results, regardless of the structure and inter-layer degree correlations of multiplex networks.   
	
	\section{Conclusions}\label{sec:conclusion}
	
	Consider the problem of rumors flooding on expanding online social networks; we initially proposed a family of containing strategies with different blocking preferences to contain the rumors spreading on correlated multiplex networks. Then we developed a heterogeneous edge-based compartmental theory to comprehend the containing dynamics. Furthermore, we compare the performance of the proposed containing strategy on multiplex networks with various topology structures and different inter-layer degree correlations. Our theory can well predict the rumor outbreak size and the threshold under all the cases studied. To the best of our knowledge, no previous studies have conducted systematic theoretical research of rumors containing dynamics on multiplex networks before.

	Extensive simulations and detailed theoretical analysis both demonstrated that the proposed strategy with preference to block users with large degree product $\mathcal{K}$ performs well in reducing the rumors outbreak size and enlarging the threshold on multiplex networks. The performance gets better with the increasing of degree heterogeneity and the intensity of preference on heterogeneous multiplex networks. 
	We further considered the inter-layer degree correlations $r_s$ of the multiplex networks. Results show that when $r_s=-1$ the strategy performs best, $r_s=1$ the second, and $r_s=0$ the last. 
	Moreover, we found that the strategy of blocking users with small $\mathcal{K}$ performs worse than the strategy of blocking users randomly on most multiplex networks except for uncorrelated multiplex networks with uniform degree distribution.
	Actually, blocking preferences will not affect the containing results on uncorrelated multiplex networks with uniform degree distribution.

	Rumor flooding is an urgent problem to be tackled. Particularly, the theoretical study of rumor containing dynamics on multiplex networks should be emphasized, as people are more and more addicted to multiple online social networks. The systematic theoretical research in this study offers inspirations for further investigations on this issue.

	\section*{Acknowledgements}
	
	This work was partially supported by the China Postdoctoral Science Foundation (Grant No.~2018M631073), China Postdoctoral Science Special Foundation (Grant No.~2019T120829), National Natural Science Foundation of China (Grant Nos.~61903266 and 61603074), and Fundamental Research Funds for the Central Universities.
	
	\appendix

	\section{}\label{nx}

	\begin{center}
		\textbf{Detailed calculations of $n_x(\vec{k},t)$}
	\end{center}

	This section is to present the detailed calculations of $n_x(\vec{k},t)$, which is the average number of neighbors in the S or R state that a spreader individual might have in either layer A or B.
	To begin with, we obtain the probability that a randomly selected edge in layer $x$ will connect to a node in the state D and I as
	\begin{equation}\label{eq:rhoI}
	f_{x}^D=1-q,
	\end{equation}
	and
	\begin{equation}\label{eq:rhoI}
	f_{x}^I=\frac{1}{\langle k\rangle_x}\sum k_x P(\vec{k})\zeta_{x}(t)^{k_{x}}\zeta_{y}(t)^{k_{y}},
	\end{equation}
	respectively.
	Nodes can only be in one of the four states; thus, a randomly selected edge in layer $x$ will connect to a node in the state S or R with a probability of $f_{x}^{SR}=1-f_{x}^I-f_{x}^D$. Then with the assumption that $u_{x}$ is in the cavity state, we get 
	\begin{eqnarray}\label{eq:neighbor}
	n_x(\vec{k},t)&=&1+g_{x}^{x}[(k_{x}-2)f_{x}^{SR}+k_{y}f_{y}^{SR}]\nonumber \\
	& &+g_{x}^{y}[(k_{y}-1)f_{y}^{SR}+(k_{x}-1)f_{x}^{SR}]
	\end{eqnarray}
	where 
	\begin{equation}
	g_{x}^{x}=\frac{(k_{x}-1)f_{x}^{SR}}{(k_{x}-1)f_{x}^{SR}+k_{y}f_{y}^{SR}}
	\end{equation}
	and
	\begin{equation}
	g_{x}^{y}=\frac{k_{y}f_{y}^{SR}}{(k_{x}-1)f_{x}^{SR}+k_{y}f_{y}^{SR}}.
	\end{equation}
	In Eq. (\ref{eq:neighbor}), the first term on the right-hand side represents the spreader node that has at least one neighboring node $w$ in the S or R state, since the spreader node can only get rumor from neighbors. The neighboring node $w$ can come from either layer $x$ or $y$; thus, we introduce $g_{x}^{x}$ and $g_{x}^{y}$ in Eq. (\ref{eq:neighbor}), of which the values reflect the probability that $w$ comes from layer $x$ or $y$, respectively.
	If $w$ comes from layer $x$, then the spreader node may have an average number of $(k_{x}-2)f_{x}^{SR}$ and $k_{y}f_{y}^{SR}$ neighbors in the S or R state in layer $x$ and $y$, respectively. Otherwise, the average number should be $(k_{x}-1)f_{x}^{SR}$ and $(k_{y}-1)f_{y}^{SR}$ in layer $x$ and $y$, respectively. 
	
	\section*{References}
	\providecommand{\newblock}{}

\end{document}